\documentclass[iop]{emulateapj}

\shorttitle{CEMP RR Lyrae Stars}
\shortauthors{Kennedy, C. R. et al.}

\newcommand{\metal}{[Fe/{}H]}
\newcommand{\cfe}{[C/{}Fe]}

\newcommand{\teff}{$T_{\rm eff}$\,}
\newcommand{\logg}{log\,$g$\,}

\begin{document}

\title{Seven New Carbon-Enhanced Metal-Poor RR Lyrae Stars}

\author{Catherine R. Kennedy}
\affil{Research School of Astronomy and Astrophysics, Australian National University, Canberra, ACT 2611, Australia}
\email{catherine.kennedy@anu.edu.au}

\author{Richard J. Stancliffe}
\affil{Argelander-Institut f\"ur Astronomie, Auf dem H\"ugel 71, 53121 Bonn, Germany}
\affil{Research School of Astronomy and Astrophysics, Australian National University, Canberra, ACT 2611, Australia}

\author{Charles Kuehn}
\affil{Sydney Institute for Astronomy, University of Sydney, Sydney, Australia}

\author{Timothy C. Beers}
\affil{National Optical Astronomy Observatory, Tucson, 
AZ 85719, USA\\ and JINA: Joint Institute for Nuclear Astrophysics}

\author{T. D. Kinman}
\affil{National Optical Astronomy Observatory, Tucson, AZ 85719, USA}

\author{Vinicius M. Placco}
\affil{Gemini Observatory, Hilo, HI 96720, USA}

\author{Henrique Reggiani, Silvia Rossi}
\affil{Departamento de Astronomia - Instituto de Astronomia, 
Geof\'isica e Ci\^encias Atmosf\'ericas, Universidade de S\~ao Paulo, 
S\~ao Paulo, SP 05508-900, Brazil}

\author{Young Sun Lee}
\affil{Department of Astronomy, New Mexico State University, Las Cruces, NM 88003, USA}

\begin{abstract}

We report estimated carbon-abundance ratios, \cfe{}, for seven newly-discovered 
carbon-enhanced metal-poor (CEMP) RR Lyrae stars.
These are well-studied RRab stars that had previously been selected
as CEMP candidates based on low-resolution spectra. For this pilot
study, we observed eight of these CEMP RR Lyrae candidates with the Wide
Field Spectrograph (WiFeS) on the ANU 2.3m telescope. Prior to this
study, only two CEMP RR Lyrae stars had been discovered: TY Gru and
SDSS~J1707$+$58.  We compare our abundances to new theoretical
models of the evolution of low-mass stars in binary systems. These
simulations evolve the secondary stars, post accretion from an AGB donor, all the way to the RR
Lyrae stage.  The abundances of CEMP RR Lyrae stars can be used as
direct probes of the nature of the donor star, such as its mass, and the amount
of material accreted onto the secondary. We find that the
majority of the sample of CEMP RR Lyrae stars is consistent with AGB
donor masses of around $1.5-2.0$ M$_\sun$ and accretion masses of a few
hundredths of a solar mass. Future high-resolution studies of these
newly-discovered CEMP RR Lyrae stars will help disentangle the effects
of the proposed mixing processes that occur in such objects.  

\end{abstract}


\section{Introduction}

Over the past few decades, it has become clear that a 
large fraction of stars with significant carbon enhancements
exists among the populations of metal-poor stars in the Galactic halo.
These carbon-enhanced metal-poor (CEMP) stars were originally defined as having
metallicities \metal{} $\le-1.0$ and carbon-abundance ratios \cfe{} $\ge+1.0$
\citep{beers2005}.  Recent studies of large numbers of metal-poor stars suggest that a more natural dividing line between the carbon-normal and carbon-enhanced populations is \cfe{} $\ge+0.7$ (see, for example, Figure 4 from Aoki et al. 2007 and Figure 4 from Carollo et al. 2012).  We therefore define CEMP stars as having \metal{} $\le-1.0$ and \cfe{} $\ge+0.7$.  

The frequency of CEMP stars, among metal-poor stars in the Milky Way,
increases with decreasing metallicity \citep{beers1992,norris1997,beers2005,cohen2005,marsteller2005,rossi2005,frebel2006,lucatello2006,
norris2007,carollo2012,lee2013,norris2013,spite2013}, as well as with
distance from the Galactic plane \citep{frebel2006,carollo2012,lee2013}. 
From the study of their elemental abundance patterns, one can begin to
uncover details concerning the nature of their progenitor objects.

There exist a number of different sub-classes of CEMP stars with
specific abundance characteristics; these different sub-classes are
suggestive of different sites of carbon production at early times
\citep{beers2005}. CEMP-$s$ stars, which exhibit evidence of
$s$-process-element enhancement, are the most common; around 80\% of
CEMP stars exhibit $s$-process-element enhancements \citep{aoki2007},
including both the CEMP-$s$ and CEMP-$r/s$ sub-classes (the latter
sub-class indicates stars for which the presence of both $r$- and
$s$-process element enhancements are found). It is widely believed that
these objects are the result of mass transfer from a companion
asymptotic giant-branch (AGB) star, where the production of carbon and
$s$-process elements occurs \citep{herwig2005,sneden2008}. About half of
the CEMP-$s$ stars have been shown to be CEMP-$r/s$, suggesting
formation from molecular clouds that had already been enhanced in
$r$-process elements, opening the possibility that their carbon
enhancements arose from more than one site. In any case, observational
evidence suggests that the CEMP-$r/s$ stars (and other
$r$-process-element rich stars) do not require a contribution of
$r$-process elements from a binary companion (see Hansen et al. 2011b,
2013). Stars in the CEMP-no sub-class exhibit no neutron-capture element enhancements, and the source of the carbon enhancement for these
stars is less certain. Current suggestions include the possibility that
very massive, rapidly-rotating, mega metal-poor (\metal{} $< -6.0$)
stars were very efficient producers of carbon, nitrogen, and oxygen, due
to distinctive internal burning and mixing episodes followed by strong
mass loss (Hirschi et al. 2006; Meynet et al. 2006, 2010). Another
suggested origin is pollution of the interstellar medium (ISM) by so-called faint supernovae
associated with the first generations of stars, which can experience
extensive mixing and fallback during their explosions \citep{umeda2003,tominaga2007,kobayashi2011,ito2013,nomoto2013,Tominaga2013}.

Recent observations of some of the lowest-metallicity RR Lyrae stars
have revealed similarities between the abundance patterns of these stars
and their non-varying counterparts in the halo of the Milky Way. For
example, Hansen et al. (2011a) find that the elemental abundance patterns
of two very metal-poor RR Lyrae stars (\metal{} $\sim-2.8$) match typical
patterns of very metal-poor stars observed in the Galactic halo, and
indeed, they can be compared to theoretical yields from the first
generations of stars.

Because a large fraction of the very metal-poor (VMP; [Fe/H] $< -2.0$)
stars studied in the Galactic halo are CEMP stars, one would expect that
a similar fraction of VMP horizontal-branch stars would exhibit carbon
over-abundances. Prior to our observational program, only two CEMP RR
Lyrae stars had been recognized \citep{preston2006,kinman2012}. Now, the
sample size has increased by more than a factor of four, as we have
identified seven new CEMP RR Lyrae stars.   

One clear advantage of the study of CEMP RR Lyrae stars, as opposed to
CEMP stars in other evolutionary stages, is that the observed surface
abundances are primarily influenced by the dilution of accreted material
in the receiving star at first dredge up. When the receiving star is on
the main sequence, it is unknown whether the observed surface
composition is purely due to the makeup of the accreted material, or
whether some non-convective process (such as rotation, e.g., Masseron et
al. 2012, or thermohaline mixing, e.g., Stancliffe et al. 2007) has
already led to dilution. By using more evolved objects we bypass this
uncertainty. Detailed stellar models \citep{stancliffe2013} suggest that
the surface composition of a CEMP star in the RR Lyrae phase is
predominantly determined by the mass of material accreted, and the
composition of the ejecta.  This assumes that all evolved objects had a mass of around 0.8 M$_\sun$ at the main sequence turn-off, and consequently, their structures along the giant branch (particularly the depth that the convective envelope reaches during first dredge-up) are similar.  The CEMP RR Lyrae stars may thus provide a
constraint on the efficiency of wind accretion, which, although currently
highly uncertain (Abate et al. 2013), potentially has an important
impact on, e.g., Type Ia supernova progenitors.

This paper is organized as follows. Section 2 describes the target
selection, observing techniques, and details of the observations.
Sections 3 and 4 describe the estimation of stellar parameters and
carbon abundances. A discussion of the Oosterhoff classifications of our program stars, and
their implications, is given in Section 5. Section 6 includes a
description of the theoretical models used for comparison, and Section 7
contains a thorough discussion of the analysis of our measured \cfe{}
abundances in terms of these theoretical scenarios. Our conclusions and
plans for future work are provided in Section 8.  

\section{Target Selection and Observations}

The RR Lyrae stars observed for this pilot program, listed in
\citet{kinman2012}, were selected from the Hamburg/ESO objective-prism
survey, cross-matched with previously recognized RR Lyraes. The benefit
of drawing CEMP candidates from the HES is that (rather crude)
preliminary estimates of \metal{} and \cfe{} from their low-resolution
spectra are available \citep{christlieb2008}. Higher-resolution
spectroscopic observations remain necessary to confirm these values, as
the errors on these low-resolution estimates are quite large.
Furthermore, observations of these RR Lyraes during the HES is at random
phase, whereas in order to obtain accurate estimates of their
atmospheric parameters (and elemental abundances) it is crucial to
obtain their spectra during times of low activity. 

\begin{deluxetable*}{lccccccccc}

  \tablecaption{Details of Targets and Observations\label{obs}}
 \tablewidth{0pc}
  \tablehead{\colhead{Star}&\colhead{R.A.}&\colhead{Decl.}&\colhead{Max Mag.}&\colhead{Amplitude}&\colhead{Waveband}&\colhead{Period}&\colhead{Ephemeris}&\colhead{Phase}&\colhead{Exposure}\\
\colhead{}&\colhead{(J2000)}&\colhead{(J2000)}&\colhead{}&\colhead{(mag)}&\colhead{}&\colhead{(days)}&\colhead{Source}&\colhead{}&\colhead{(sec)}}
\startdata
IV Leo       & 10:58:12.6 & -00:05:40 & 15.4 & 0.4&$V$& 0.6358&3&  0.436 & 1200 \\
LO Leo       & 11:35:22.7 & -00:53:42 & 14.8 & 1.1&$V$& 0.6069&3 & 0.510& 1800\\
LP Leo       & 11:36:39.8 & -01:25:16 & 16.5 & 0.5&$V$& 0.6492&3 & 0.682& 2400\\
v370 Vir     & 12:06:04.1 & -02:12:57 & 14.7 & 0.9&$V$& 0.6992&3 & 0.455& 1800\\
v408 Vir     & 12:40:03.5 & -00:04:09 & 16.5 & 0.7&$V$& 0.5858&3 & 0.391& 3600\\
J1245$-$0419 & 12:45:03.0 & -04:19:11 & 16.1 & 0.4&$V$& 0.7458&2 & 0.608& 3600\\
ZZ Vir       & 13:23:38.6 & -04:21:42 & 13.7 & 1.2&$B$& 0.6841&1& 0.614& 600 \\
WY Vir       & 13:35:16.1 & -06:58:22 & 13.0& 1.2&$B$& 0.6093&1 & 0.769& 300\\
\enddata
\tablerefs{(1) Samus et al. (2009); (2) Miceli et al. (2008); (3) Vivas et al. (2004).}
\end{deluxetable*}

The variable nature of RR Lyrae atmospheres poses some challenges for
carrying out spectroscopic observations. The radial pulsations of these
stars means that their surface gravity and effective temperature will
vary significantly over short intervals of time. These pulsations also
impact the star's spectral lines, causing them to shift due to the
radial velocities of the gas in the outer layers of the star; variations
of up to 70 km s$^{-1}$ are found during a pulsation cycle (Smith 2004).  Ideally, RR
Lyrae stars should be observed when their phase is in the range
$\phi=0.2-0.8$, corresponding to minimum light; maximum light occurs at
$\phi=0$ and $\phi=1$.

All of the stars we observed have had their periods determined
previously, allowing us to generate the accurate ephemerides
required to determine the phases.  All ephermerides were adjusted to agree with light curves derived from the data available in January 2013 from the Catalina Surveys Data Release 2 (CSDR2)\footnote{http://crts.caltech.edu/index.html}.  The variability of RR Lyraes also
constrains the length of the exposure times for the spectroscopic
observations, as observing for too long could result in artificial
broadening of the lines due to the changes in the atmospheric properties
and radial velocities that are part of the pulsation cycle. For our
targets, which are all RRab stars, we noted that the exposure times
around any given phase should be less than 1.5 hours.  

Observations of the first eight CEMP candidates were obtained in March 2013 using the
Wide Field Spectrograph \citep[WiFes;][]{dopita2007} on the ANU
2.3-meter telescope at Siding Spring Observatory. The gratings used were
both the B3000 and R7000, producing resolving powers of $R\sim3000$ and
7000, respectively. The B3000 grating was deliberately selected because
the wavelength coverage (3500$-$5700 \AA{}) and resolution for this set up are
directly compatible with the automated pipeline employed for atmospheric
parameter determination, which is further described in Section
3.   These data were reduced
with the PyWiFeS software package \citep{childress2013}.

The details of our observed RRab stars, including coordinates, maximum magnitude and amplitude for a given waveband, and the source of ephemeris for each star, are given in Table \ref{obs}.  We also list the exposure time and phase at mid-observation for each observed object.

\section{Atmospheric Parameters}

The atmospheric parameters were determined with the n-SSPP, a version of
the original SEGUE Stellar Parameter Pipeline (SSPP; Allende Prieto et
al. 2008; Lee et al. 2008a, 2008b, 2011; Smolinski et al. 2011) that is
capable of working with non-SEGUE spectroscopic data and without Sloan
colors available. Given input moderate-resolution spectra and color
information for a number of photometric systems, the pipeline determines values for \teff{}, \logg{}, and \metal{} by averaging over a series of estimates
for each parameter, based on spectral synthesis and $\chi^{2}$
minimization using grids of synthetic spectra, as well as other available
techniques. In the case of v370 Vir, the \logg{} was set at a value of 2.5, as the
estimates from the n-SSPP proved to be unreliable for this spectrum. We
note that, for RR Lyrae stars in these observed phases, the value of
\logg{} should be less than 3.0 \citep{barcza2009}. We adopt the
nominal uncertainties of 150 K for \teff{}, 0.5 dex for \logg{}, and 0.3
dex for \metal{}. For each RR Lyrae star, we set the microturbulent
velocity at $\xi=3$ km s$^{-1}$, and adopt a conservative uncertainty of
1 km s$^{-1}$.  

\begin{figure}[!ht]
\epsscale{1.15}
\plotone{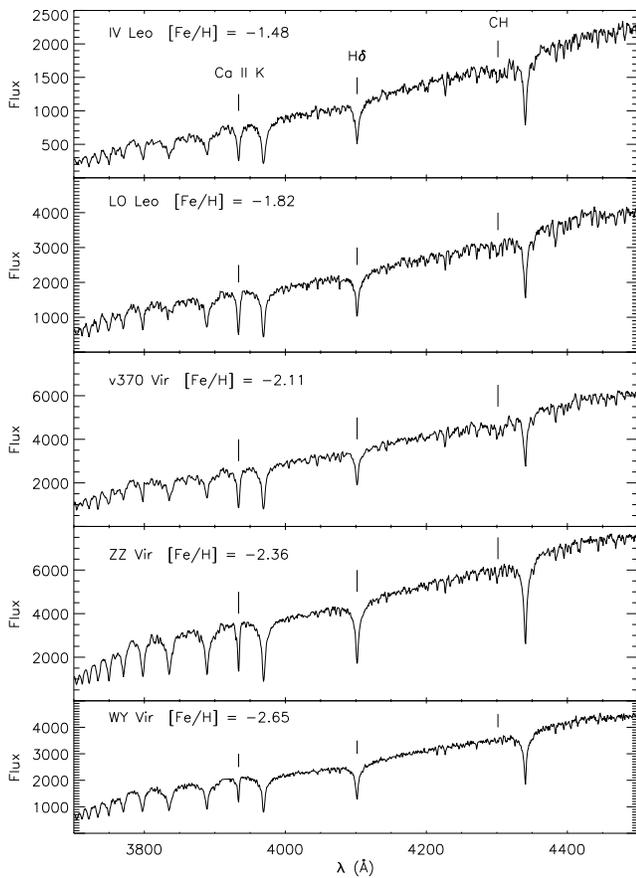}
\caption{Moderate-resolution WiFeS spectra for five program stars in 
order of decreasing metallicity. The Ca {\sc{ii}} K line, the H$\delta$
Balmer line, and the CH $G$-band for each star are indicated by the
vertical lines.}
\label{spectra}
\end{figure}

Examples of the WiFeS moderate-resolution spectra for five of our
program stars are shown in Figure \ref{spectra}, in order of decreasing
metallicity. Note that one cannot directly compare the relative
strengths of the Ca {\sc{ii}} K lines because of differences in the effective temperatures of these stars.

\section{Carbon-Abundance Ratios}

Carbon-abundance ratios, \cfe{}, are determined by spectral synthesis of the
CH $G$-band at 4300 \AA. For this analysis, we employ the
one-dimensional plane-parallel stellar atmospheres of
\citet{castelli2004}, with no convective overshoot. Solar abundances
used for reference are those of \citet{asplund2009}. The 2013 version of
MOOG \citep{sneden1973} was employed for spectral synthesis, and we
minimized $\chi^{2}$ over the region 4290$-$4320 \AA{} for each
spectrum.   

\begin{deluxetable*}{lcccccccc}
%\tabletypesize{\scriptsize}

  \tablecaption{Atmospheric Parameters and C Abundances\label{results}}
 \tablewidth{0pc}
  \tablehead{\colhead{} & \colhead{} & \multicolumn{4}{c}{This Study}& \multicolumn{3}{c}{Previous Estimates}\\
\cline{7-9}\\
\colhead{} & \colhead{} & \colhead{} & \colhead{} & \colhead{}&\colhead{}&\multicolumn{2}{c}{Christlieb\tablenotemark{b}} & \colhead{Other}\\%
%\cline{3-6}
\colhead{Star}&\colhead{$\Delta$log$P$\tablenotemark{a}} & \colhead{\teff{} (K)}&\colhead{\logg{} (cgs)}& \colhead{\metal{}}&\colhead{\cfe{}}& \colhead{\metal{}}&\colhead{\cfe{}}&\colhead{\metal{}}} 

\startdata
IV Leo & $+$0.0108& 6609 & 2.82 & $-$1.48 &  $+$1.09 (0.20) & $-$2.8 & $+$0.2 & \nodata\\
LO Leo & $+$0.0606& 6394 & 2.85 & $-$1.82 & $+$0.73 (0.25) & $-$2.7 & $+$1.1 & $-$1.71\tablenotemark{c}\\
LP Leo & $-$0.0122& 5882 & 2.61 & $-$1.64 & $+$0.66 (0.20) & $-$1.7 & $-$0.4 & \nodata\\
v370 Vir & $+$0.0833 & 6275 & 2.50 & $-$2.11 & $+$1.36 (0.20) & $-$2.8 & $+$1.7 & $-$2.08\tablenotemark{c}\\
v408 Vir & $-$0.0177 & 6216 & 2.81 & $-$1.88 & $+$0.62 (0.30) & $-$1.4 & $+$0.7 & \nodata\\
J1245$-$0419 & $+$0.0535& 5996 & 2.76 & $-$1.87 & $+$0.34 (0.30)& \nodata & $+$2.4 & \nodata\\
ZZ Vir & $+$0.0736&  6538 & 2.54 & $-$2.36 & $+$1.71 (0.25) & $-$2.9 & $+$0.9 & \nodata\\
WY Vir & $+$0.0603& 6354 & 2.81 & $-$2.65 & $+$1.35 (0.30) & \nodata & $+$3.9 & $-$2.80\tablenotemark{d}\\
\enddata
\tablenotetext{a}{Shift in log period from the locus of Oo I stars in Bailey diagram.}
\tablenotetext{b}{Low-resolution estimates of metallicities and carbon-abundance ratios from \citet{christlieb2008}.}
\tablenotetext{c}{Vivas et al. (2008).}
\tablenotetext{d}{Layden 1995.}
\end{deluxetable*}

\begin{figure}[!ht]
\epsscale{1.15}
\plotone{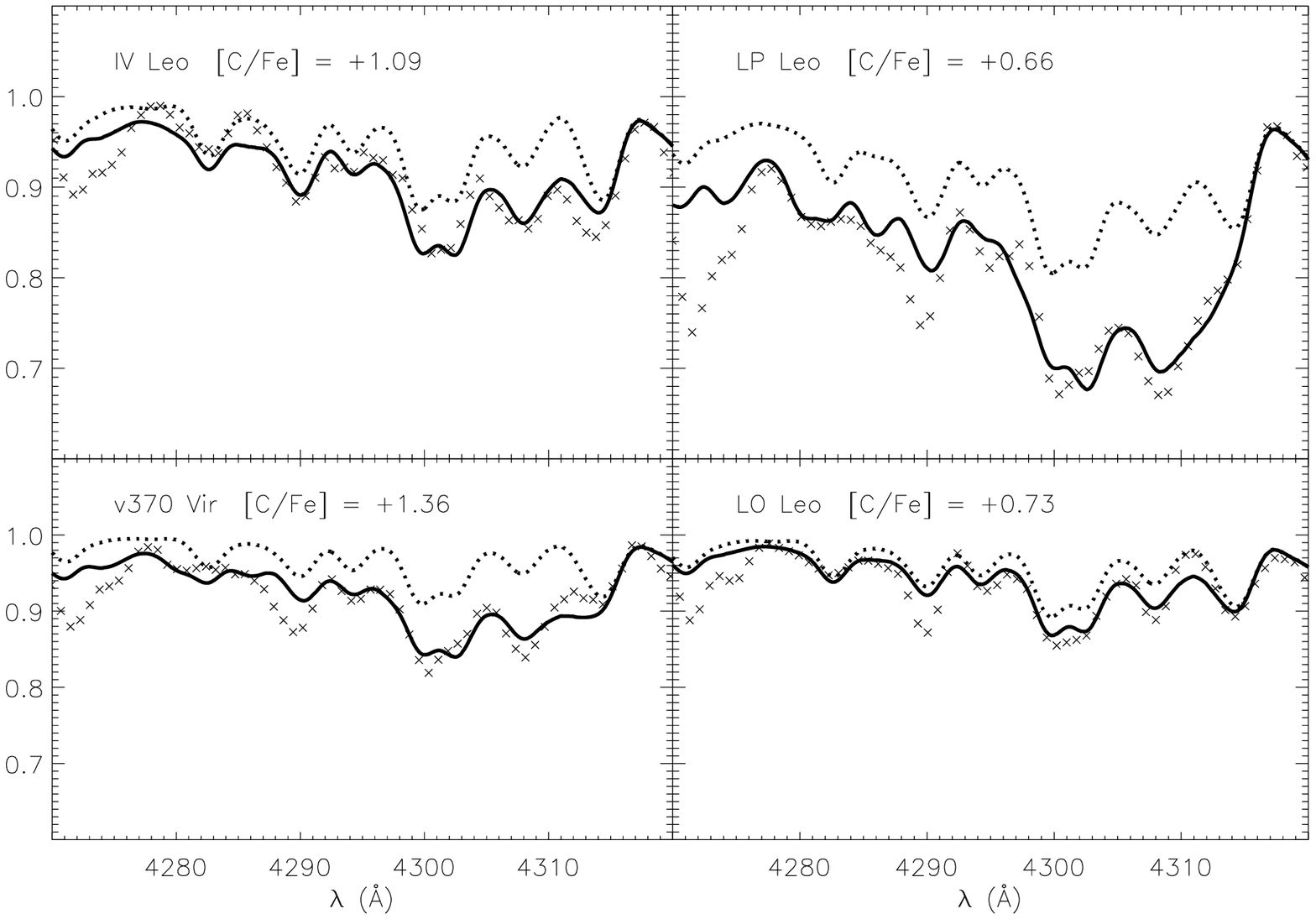}
\caption{Synthesis of the CH $G$-band for four stars in the sample.  Crosses represent the observed spectrum for each object.  The solid lines are the best-fitting synthetic spectra, while the dotted
lines show the solar value, [C/Fe] = 0.}
\label{cfeall}
\end{figure}

\begin{figure}[!ht]
\epsscale{1.15}
\plotone{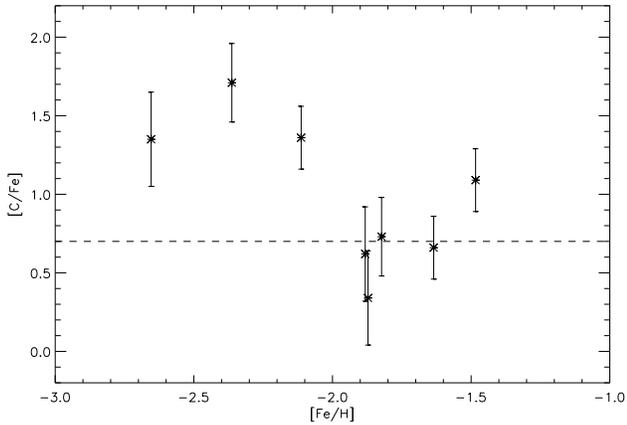}
\caption{\cfe{} vs. \metal{} for the entire program sample.  Stars with
[C/Fe] values falling above the dashed line at \cfe{} $=+0.7$ are classified as CEMP
stars. There are two stars that lie very close to the line, while another is
consistent with it to within the error bars. We consider all but one
of these stars to be CEMP stars.}
\label{cfefeh}
\end{figure}

Examples of this procedure can be seen in Figure \ref{cfeall}, where we
show the synthesis of the CH $G$-band for four of our eight program
stars. For each of the panels, the solid line denotes the best-fitting
synthetic spectrum, corresponding to the best estimate of \cfe{} for
each star. For reference, the dotted line denotes the solar value of the
carbon-abundance rato, \cfe{} $=$ 0. The atmospheric parameters and carbon
abundances for all eight RR Lyrae stars in the sample are listed in
Table \ref{results}.  Figure \ref{cfefeh} shows the \cfe{} ratios, as a
function of [Fe/H], for our program sample. There are seven stars which, at
least within error bars, pass the criterion \cfe{} $\ge+0.7$ for
inclusion as CEMP stars.

 We also include in Table \ref{results} the low-resolution estimates of \metal{} and \cfe{} from \citet{christlieb2008}, used for target selection, as well the \metal{} estimates from other studies of RR Lyrae stars \citep{vivas2008,layden1995}, available for three stars in our sample.  Because the \citet{christlieb2008} calculation of \cfe{}  is based on the CH $G$-band strength at a given color, it does not require an
estimate of \metal{}.  If the \ion{Ca}{2} K line was too weak (on the prism spectrum) to be detected, an estimate of \metal{} could not be obtained.  Even with the aforementioned large errors on \metal{} and \cfe{} from the low-resolution prism spectra, we note that five out of six stars with estimates of \cfe{} $\ge +0.7$ from \citet{christlieb2008} turned out to be CEMP stars.  Furthermore, the additional estimates of metallicity for LO Leo, v370 Vir, and WY Vir agree with our \metal{} estimates to within $0.15$ dex.

\subsection{Uncertainties}

\begin{figure}[!ht]
\epsscale{1.15}
\plotone{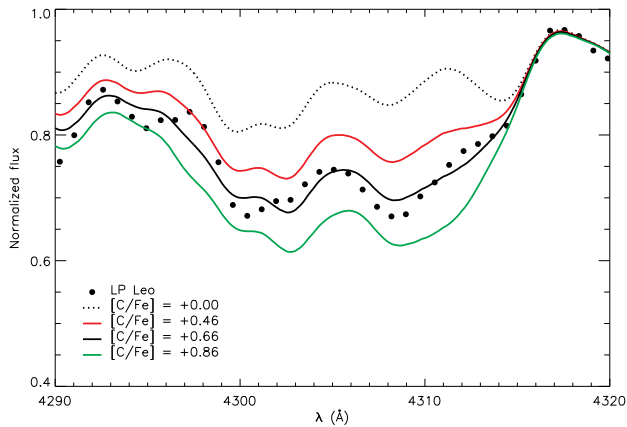}
\caption{Example of \cfe{} uncertainty determination.  The solid line represents
the best-fitting spectrum, while the red and green lines represent the synthetic
spectra associated with our estimated error of $\delta$\cfe{}$=0.2$ dex for LP Leo.  
Again, the dotted line show the solar value, [C/Fe] = 0.}
\label{cfefit}
\end{figure}

Uncertainties for the carbon-abundance ratios, $\delta$\cfe{}, were determined
directly from the spectral synthesis. We estimated the upper and lower
bounds by demanding that their associated synthetic spectra completely
enclose the CH $G-$band feature from which we estimate \cfe{}. An example of this
technique is shown in Figure \ref{cfefit}. The uncertainty in
\cfe{} for each star is shown in parentheses next to its carbon-abundance ratio in Table \ref{results}. The error on \cfe{} for this sample
ranges from 0.2$-$0.3 dex.   

We also have estimated the systematic uncertainties, those due to
adopted errors on our atmospheric parameters, for each star in the
sample. The average values of these systematic uncertainties are listed
in Table \ref{uncert}. It is clear that the two largest possible
systematic uncertainties arise from errors associated with effective
temperature and metallicity. Even very conservative adopted
uncertainties for surface gravity and microturbulent velocity do not
result in the propagation of significant errors on our estimated \cfe{}
abundances.

\begin{deluxetable}{lc}

%\tabletypesize{\scriptsize}

  \tablecaption{Systematic \cfe{} Uncertainties\label{uncert}}
 \tablewidth{.25\textwidth}
  \tablehead{\colhead{}&\colhead{$\Delta$ \cfe{}}}
%\colhead{}&\colhead{(K)}&\colhead{(cgs)}&\colhead{}&\colhead{}}
\startdata
\teff{}$-$150 K & $-$0.30\\
\teff{}$+$150 K & $+$0.28\\
\logg{}$-$0.5 dex & $+$0.16\\
\logg{}$+$0.5 dex & $-$0.16\\
\metal{}$-$0.3 dex & $+$0.36\\
\metal{}$+$0.3 dex & $-$0.43\\
$\xi$$-$1 km s$^{-1}$ & $+$0.05\\
$\xi$$+$1 km s$^{-1}$ & $-$0.06\\
\enddata

\end{deluxetable}

\section{Oosterhoff Classification}

Milky Way globular clusters can be divided into two groups, known as Oosterhoff groups, based on the properties of their RR Lyrae stars, with Oosterhoff II (Oo II) clusters being more metal poor and having longer average RR Lyrae periods than Oosterhoff I (Oo I) clusters \citep{oosterhoff1939}.  Like the RR Lyrae variables studied in globular clusters, RR Lyrae field stars have been shown to exhibit properties of both Oo I and Oo II systems \citep{kinemuchi2006}, although the separation between the two populations is far less distinct.  Following \citet{kinman2012}, we study the Oosterhoff types of the CEMP RR Lyrae field stars in order to compare to classical Oosterhoff systems, as well as to explore how the presence of carbon in the atmosphere might affect the nature of the pulsation.   

\begin{figure}[!ht]
\epsscale{1.15}
\plotone{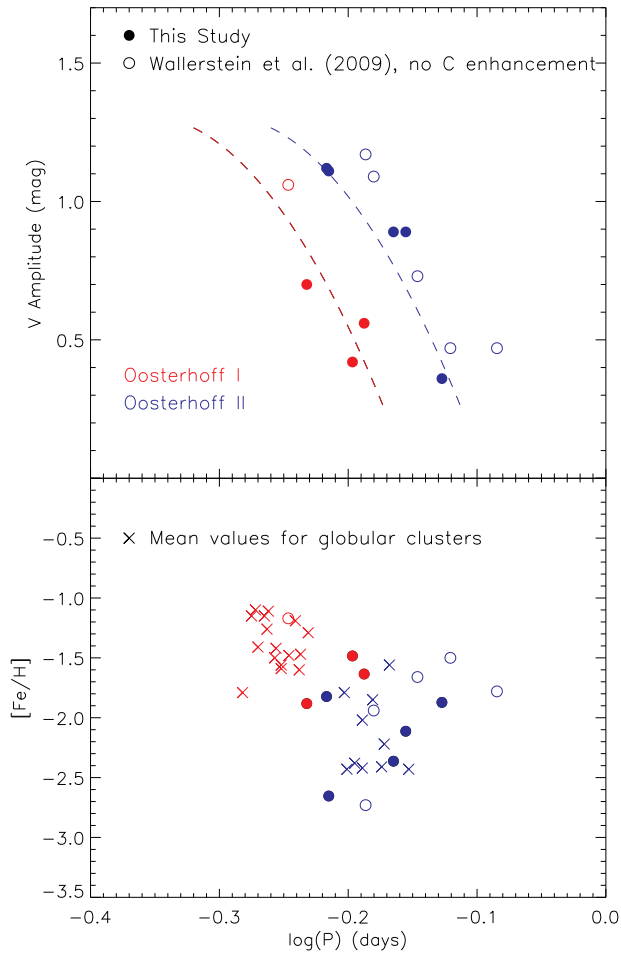}
\caption{Oosterhoff classification of CEMP RR Lyrae stars.  {\it Top panel:}  Period-amplitude diagram
for program stars (filled circles) and those from \citet{wallerstein2009} (open circles).  
Oo I (red) and Oo II (blue) types are assigned based on their proximity to the trends of \citet{cacciari2005} (dashed lines).
{\it Bottom panel:}  Metallicity vs. period for RR Lyrae stars (same symbols as above) 
compared to the {\it mean} values found in studies of globular clusters \citep{bono2007}.}
\label{ooster}
\end{figure}

The Oosterhoff types of the observed CEMP RR Lyrae stars are determined
based on their location in the period-amplitude plot (or Bailey
diagram), as shown in the top panel of Figure \ref{ooster}. The red and
blue dashed lines show the typical loci for Oo I and Oo II RRab stars
\citep{cacciari2005}.  More quantitatively, stars with $\Delta$log$P$ $>$ 0.50 are classified as Oosterhoff II, where $\Delta$log$P$ is the shift in the log period from the locus of Oo I stars at a given amplitude.  The values of $\Delta$log$P$ for each object as given in the second column of Table \ref{results}.  Based on the periods and $V$-band amplitudes of
the sample, we classify three stars as Oo I types and the remaining five as
Oo II types (solid circles). Also shown in this figure are the periods
and amplitudes for a sample of RR Lyrae stars with \metal{} $\le-1.0$
from \citet{wallerstein2009} (open circles). Unlike our sample, these
metal-poor RR Lyrae stars do not exhibit carbon enhancements. In fact,
the majority of them have carbon-abundance ratios that are significantly
sub-solar (\cfe{} $\sim-0.5$). These lower carbon abundances can perhaps
be attributed to CN cycling having occurred as these stars ascended the
giant branch. In this case, the stars would likely have originally been
carbon-normal metal-poor stars, with \cfe{} $\sim+0.0$, prior to having
undergone evolutionary mixing effects (see, e.g., Figure 3 in Lucatello et al. 2006).  

In the lower panel of Figure \ref{ooster}, we compare the metallicities
and periods of the Oo I and Oo II RR Lyrae stars from above to the mean
values found in globular clusters (shown as crosses), taken from
\citet{bono2007}.  These globular cluster classifications are based
on an average of many RRab stars within each cluster, and there is an
appreciable range in pulsation periods among the individual objects.
Thus, a direct comparison of field RR Lyrae stars to mean values for
globular clusters is perhaps not meaningful. Nevertheless, we recognize
that our observed Oo I and Oo II field stars are consistent with the
general behavior of Oo I and Oo II clusters.  

Given that there are very few RR Lyrae stars with available carbon
abundances, it is difficult to fully assess how the presence of enhanced
carbon in the atmosphere might affect the physical nature of these
stars. The carbon-weak field RR
Lyrae stars from \citet{wallerstein2009} appear to exhibit longer
pulsation periods than their carbon-enhanced counterparts, even though the metallicities of the two samples are very
similar.  However, the periods of the carbon-weak stars are still within the range of those periods associated with Oo II stars, and their separation from the Oo II locus is within the level of scatter that is seen in Oo II clusters.  We therefore are unable to conclude that the presence of enhanced carbon influences the dynamics of pulsation.  Nevertheless, we seek to obtain carbon-abundance information for a much larger sample of RR Lyrae stars, since the contrast in the behavior
of pulsation between carbon-enhanced and carbon-weak samples could be indicative that the
presence (or lack) of significant carbon has a physical effect on the
dynamics of pulsation or other properties of horizontal-branch stars.  

\section{Theoretical Models of CEMP RR Lyrae Stars}

Recently, \citet{stancliffe2013} conducted thorough simulations of the
binary progenitor systems of CEMP RR Lyrae stars with \metal{} $=-2.3$.
In these models, a range of masses of accreted material from AGB donor
stars (with initial masses of 1$-$2 M$_\sun$) are simulated, following
the prescription of \citet{stancliffe2009}. Post accretion, the
secondary star is then evolved all the way through full giant-branch
evolution to the horizontal-branch stage. Using the detailed yields for
low-metallicity AGB stars, computed by \citet{lugaro2012}, the
composition of the companion as it evolves is predicted, for any species
from H to Pb (the terminus of the $s$-process), rather than just for the
light elements (as computed by Stancliffe 2009).   We refer the interested reader to \citet{stancliffe2013} for details of the code.  Throughout these
simulations, various evolutionary mixing scenarios are considered,
including combinations of standard convection, thermohaline mixing, and
gravitational settling. The final theoretical abundances of the CEMP RR
Lyrae stars are thus the result of complete consideration of primary
mass, AGB nucleosynthesis, accretion mass, and subsequent mixing physics
(both convective and non-convective) for a particular set of initial
conditions. Therefore, these models present themselves as realistic
comparisons to our observed CEMP RR Lyrae stars. 

While the theoretical models include abundance information for several
elements, including those formed via the $s$-process, the sample of
observed CEMP RR Lyrae stars presently only have carbon-abundance
information, and therefore, we restrict our theoretical comparisons to
carbon alone. For reference, a table of the theoretical \cfe{}
abundances used for our comparison is given in Table \ref{model}. For
each combination of donor mass and accreted mass, the resultant \cfe{}
values are given under three different assumptions of mixing physics:
standard convection only (C), standard convection and thermohaline
mixing (C/TH), and standard convection, thermohaline mixing, and
gravitational settling (C/TH/GS).

\begin{deluxetable}{ccccc}

%\tabletypesize{\scriptsize}

  \tablecaption{Theoretical \cfe{} from \citet{stancliffe2013}\label{model}}
 \tablewidth{0pc}
  \tablehead{\colhead{AGB Donor Mass}&\colhead{Accreted Mass}&\multicolumn{3}{c}{[C/Fe]}\\
\cline{3-5}\\
\colhead{(M$_\sun$)}&\colhead{(M$_\sun$)}&\colhead{C}&\colhead{C/TH}&\colhead{C/TH/GS}}
%\colhead{}&\colhead{(K)}&\colhead{(cgs)}&\colhead{}&\colhead{}}
\startdata
1.0 & 0.1 & $+$0.72 & $+$0.65 & $+$0.52\\
1.0 & 0.01 & $-$0.28 & $-$0.23 & $-$0.62\\
1.0 & 0.001 & $-$0.55 & $-$0.52 & $-$0.93\\
 & & & &\\
1.5 & 0.1 & $+$2.26 & $+$2.07 & $+$2.07\\
1.5 & 0.05 & $+$1.73 & \nodata & \nodata\\
1.5 & 0.02 & $+$0.97 & \nodata & \nodata\\
1.5 & 0.01 & $+$0.52 & $+$0.48 & $+$0.39\\
1.5 & 0.001 & $-$0.58 & $-$0.44 & $-$0.57\\
 & & & &\\
2.0 & 0.1 & $+$2.58 & $+$2.02 & $+$2.04\\
2.0 & 0.01 & $+$0.95 & $+$0.62 & $+$0.61\\
2.0 & 0.001 & $-$0.52 & $-$0.36 & $-$0.42\\
\enddata

\end{deluxetable}

\section{Discussion}

\begin{figure}[!ht]
\epsscale{1.15}
\plotone{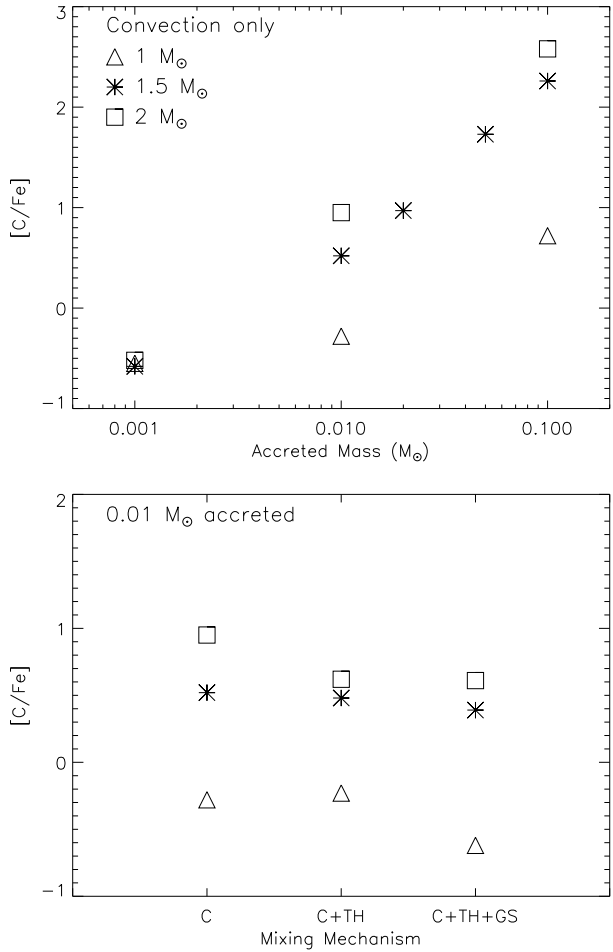}
\caption{Theoretical \cfe{} from \citet{stancliffe2013}.  {\it Top panel:}  Dependence of \cfe{} on accreted mass for a range of AGB donor masses.
Only the standard convection scenario is shown.  {\it Bottom panel:}  Dependence of \cfe{}
on mixing scenario for a range of AGB donor masses (same symbols as above)
for an accretion mass of 0.01 M$_\sun$.}   
\label{cfemodel}
\end{figure}

In comparing our \cfe{} abundances to the theoretical abundances of
\citet{stancliffe2013}, we explore the likely AGB donor masses,
accretion masses, and mixing histories associated with the progenitor
systems of our observed CEMP RR Lyrae stars. In Figure \ref{cfemodel},
the dependence of \cfe{} on donor mass, accreted mass, and mixing
mechanism is shown for some of the theoretical scenarios given in Table
\ref{model}. It is clear from the top panel of Figure \ref{cfemodel}
that the \cfe{} abundances are strongly dependent on the assumed
accreted mass, while the abundances are much less sensitive to the
assumed mixing mechanism (bottom panel). Indeed, given that the errors
on our \cfe{} estimates are around $\sim0.25$ dex, we are unable to
place any conclusive constraints on the evolutionary mixing processes
that have occurred in the observed stars. Nevertheless, the AGB donor
mass and the mass of the accreted material are constrained for the
entire sample.  

\subsection{\metal{} Considerations and \cfe{} Corrections}

All of the theoretical models available for comparison
\citep{stancliffe2013} correspond to \metal{}$=-2.3$, and include as
inputs AGB yields for \metal{}$=-2.3$ \citep{lugaro2012}. However, the
metallicities of the observed CEMP RR Lyrae stars, given in Table
\ref{results}, span a wide range ($-2.65<$ \metal{} $<-1.48$). Given
that the theoretical carbon yields of metal-poor AGB stars vary
dramatically for models of different metallicity (for reference, see
Figure 5 from Campbell \& Lattanzio 2008), it is necessary to apply a
metallicity-dependent correction to the theoretical \cfe{} abundances
prior to comparison to our observed \cfe{}.  

To determine the metallicity-dependent \cfe{} corrections, we employed
new theoretical AGB yields for \metal{}$=-1.4$ and $-1.8$ \citep{karakas2014}, in combination with the
[Fe/H] = $-2.3$ yields, for an AGB donor mass of 1.7 M$_\sun$. With the \cfe{}
values for these three different metallicities in hand, we performed a
linear interpolation with \metal{} in order to estimate the \cfe{}
correction, $\Delta${[C/Fe]}, for each observed RR Lyrae star. These
eight individual corrections were then applied to the theoretical \cfe{}
abundances in Table \ref{model}, prior to comparison with our results.
While the corrections were estimated from the direct AGB yields instead
of the surface abundances of the evolved secondary, we note that the
subsequent {\it relative} dilution of the material, once accreted,
should not vary appreciably with metallicity. A test of these adopted
corrections has been carried out for one scenario (2 M$_\sun$ donor, 0.1
M$_\sun$ accretion, \metal{}$=-1.8$), and the final, post-evolution
\cfe{} is consistent with our adopted \cfe{} based on the correction
described above.  

For the interpretation of these results, we apply the aforementioned \cfe{}
corrections directly to the models, in order to approximate the
theoretical yields for RR Lyrae stars of different metallicities.
However, for visual simplicity, we show graphically in Figure \ref{cfe}
the {\it opposite} shifts applied to the \cfe{} estimates of the
individual RR Lyrae stars. In this figure, the open squares are the
measured \cfe{} values, and the red squares show the \cfe{} values after
they have been appropriately scaled. The horizontal lines show the
theoretical \cfe{} values for different model scenarios, for
\metal{}$=-2.3$. Hence, for theoretical comparison, one should focus
only on the shifted values of \cfe{} (shown as red squares) in Figure
\ref{cfe}, as they have been shifted in order to match the metallicity
of the models, \metal{}$=-2.3$.  

\subsection{Progenitor Scenarios for Individual CEMP RR Lyrae Stars}

\begin{figure*}[!ht]
\epsscale{1.15}
\plotone{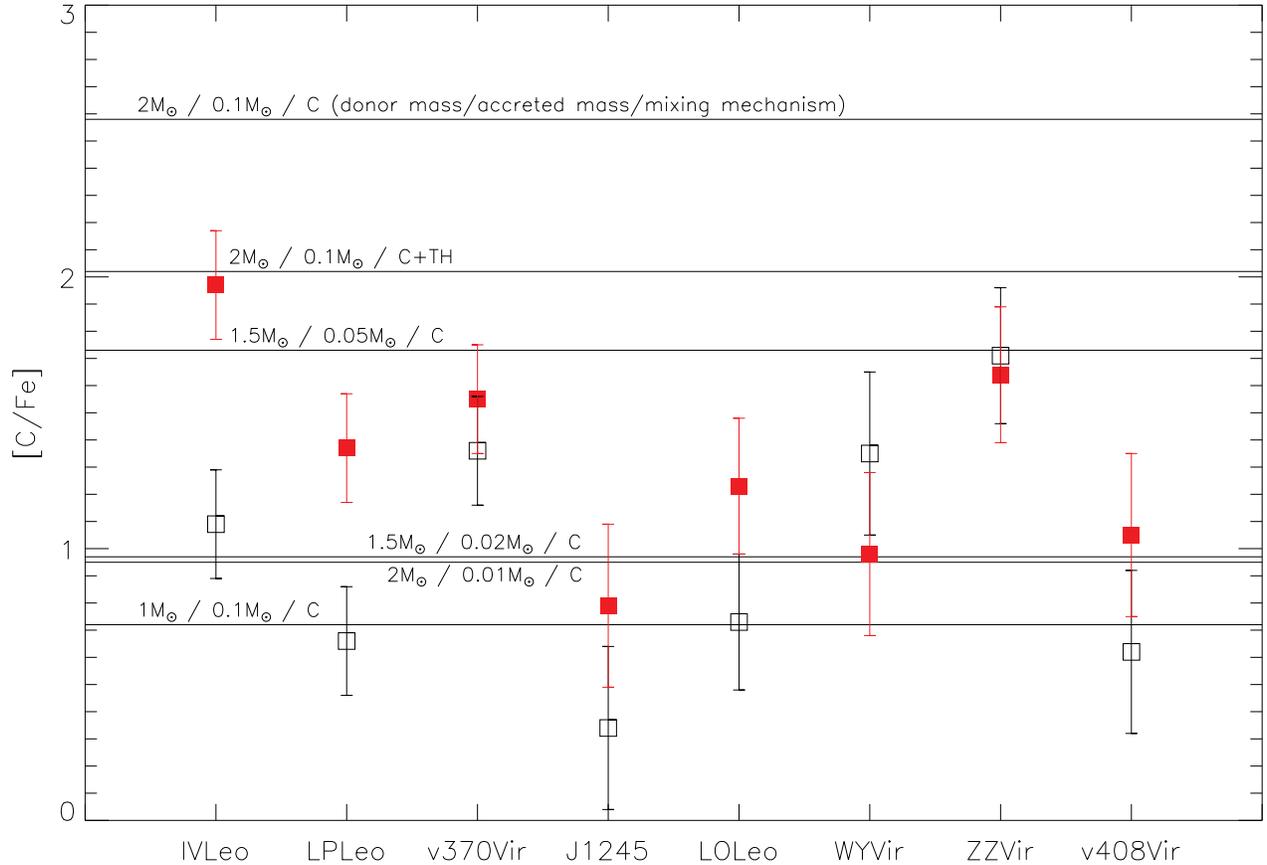}
\caption{Comparison of \cfe{} abundances in CEMP RR Lyrae stars to theoretical
models (horizontal lines).  The black open squares are the measured abundances.  
The solid red squares have been shifted to match the metallicity of the models, and therefore
they are the focus for accurate comparison.  {\it See text for details.}}
\label{cfe}
\end{figure*}

{\it LP Leo, v370 Vir, LO Leo, ZZ Vir, v408 Vir --} The \cfe{}
abundances of these five CEMP RR Lyrae stars, after the applied
correction on \cfe{} is considered, are indicative of progenitor systems
that share several common properties. In Figure \ref{cfe}, it is evident
that none of these five stars are consistent with the lowest-mass AGB
donor, 1 M$_\sun$, even when the highest accreted mass is considered.
For lower accretion masses, the predicted values of \cfe{} are sub-solar
and not shown in the figure. In fact, for all AGB donor masses
considered, the lowest accretion masses (0.001 M$_\sun$) are
inconsistent with this sample, as the resultant \cfe{} abundances are
all sub-solar. Furthermore, these observed abundances are inconsistent
with large accretion masses, 0.1 M$_\sun$, regardless of the AGB donor mass.
For these five stars, we therefore can reasonably constrain the AGB
donor mass as being $>1$ M$_\sun$ and the accreted mass as being $<0.1$
M$_\sun$. Indeed, these five CEMP RR Lyrae stars are consistent with AGB
donors in the range $\sim 1.5-2.0$ M$_\sun$ and appear to have accreted
a few hundredths of a solar mass of AGB material prior to their
subsequent evolution.

{\it WY Vir --} The lowest-metallicity star in the sample, WY Vir, has a
carbon-abundance ratio that exactly matches two of the theoretical scenarios,
both 0.02 M$_\sun$ of material accreted from a 1.5 M$_\sun$ donor and
0.01 M$_\sun$ of material accreted from a 2.0 M$_\sun$ donor, as shown
in Figure \ref{cfe}. As such, it appears to be indicative of the same
type of progenitor system described for the five stars above. However,
given the large uncertainty associated with the measured \cfe{} (0.3 dex), there remains a possibility that $\sim0.1$ solar
masses of material could have been accreted from a low-mass AGB
companion of $\sim1$ solar mass. While this ambiguity exists, it is
nevertheless evident that such large accretion masses cannot have come
from the larger-mass donors considered. And, as for the rest of the
sample, we can rule out all scenarios of very low accretion masses of
0.001 M$_\sun$.  

{\it IV Leo --} The highest-metallicity star in the sample, IV Leo,
presents the only opportunity to explore the different evolutionary
mixing mechanisms explored in these models. Similar to all of the stars
described above, its measured \cfe{} agrees with models of accretion of
a few hundredths of a solar mass from a 1.5 or 2 solar mass AGB donor.
Interestingly, another possible progenitor scenario includes the
accretion of a large amount of mass ($\sim$0.1 M$_\sun$) from the 1.5 or
2 solar mass donor, but {\it only} if one considers that thermohaline
mixing, or thermohaline mixing and gravitational settling, was/were
active during a previous evolutionary phase. Without these extra mixing
processes, such a large accretion mass would result in a much larger \cfe{} than we measure in this star. It is possible,
therefore, to consider that a very large amount of mass was accreted
under these certain mixing physics constraints. We confidently rule out
any scenario with an AGB companion of 1 solar mass, as well as all cases
of 0.001 solar mass accretion.  

{\it J1245$-$0419 --} The only star in this sample which does not
formally meet the criteria for CEMP stars, J1245$-$0419, has the most
ambiguous carbon-abundance ratio when it comes to theoretical comparison, as
can be seen in Figure \ref{cfe}. The value of \cfe{} for this star is
fully consistent with a wide range of scenarios, including all three
donor masses. If the AGB companion was only 1 M$_\sun$, then accretion
must have been very high. For the 1.5 and 2 M$_\sun$ cases, accretion
mass would have to be around 0.01 or 0.02 solar masses. Regardless of
this ambiguity, we can with certainty rule out a large accretion mass
from a 1.5 or 2 solar mass donor.  

\section{Conclusions}

We have obtained moderate-resolution spectroscopy and estimated stellar
parameters and carbon-abundance ratios, \cfe{}, for a sample of 8 RR Lyrae stars,
resulting in the identification of seven new CEMP RR Lyrae variables. As
there were previously only two such stars discovered, we significantly
increase the sample size of these objects.  

Five of the stars are Oosterhoff II type variables and the remaining
three are Oosterhoff I, based on their periods, $V$ amplitudes, and
metallicities. Upon comparison to a sample of metal-poor RR Lyrae stars
with sub-solar \cfe{}, we find a separation between pulsation
period, suggesting that the presence (or lack) of sufficient carbon
could have a physical effect on the dynamics of variability. However,
larger sample sizes of both carbon-enhanced and carbon-weak RR Lyrae
stars remain necessary to fully explore this claim.  

We compared the measured \cfe{} abundances in the CEMP RR Lyrae stars to
new theoretical models of AGB binary mass transfer and subsequent
evolution. We find that the majority of the are consistent with AGB
donor masses of $\sim1.5-2.0$ M$_\sun$ and accretion masses of
$\sim0.01-0.05$ M$_\sun$. Large accretion masses, on the order of 0.1
M$_\sun$ remain possible for a few objects. All eight stars are
inconsistent with scenarios in which very low accretion masses of
$\sim0.001$ M$_\sun$ are considered, which is to be expected, given the small contribution of carbon these donors provide and the large degree of dilution that this material undergoes.  

Future studies should include high-resolution spectroscopy of CEMP RR
Lyrae stars, as well as other non-variable horizontal-branch stars, in
order to explore more elements for theoretical comparison, in particular
those associated with the $s$-process. With a larger set of
high-precision abundance estimates, we can delve more deeply into the
likely mixing history of these stars (by both convective and
non-convective processes).  Furthermore, we plan to expand the set of
theoretical models to include those of different metallicities for a
more complete set of donor masses and accretion masses.  

\acknowledgments{CRK acknowledges support from the Australian Research Council (Super Science Fellowship; FS110200016).  RJS is the recipient of a Sofja Kovalevskaja Award from the Alexander von Humboldt Foundation.  TCB acknowledges partial support from grant PHY 08-22648; Physics Frontier Center/Joint Institute for Nuclear Astrophysics (JINA), awarded by the US National Science Foundation.  The authors thank Amanda Karakas for providing AGB yields for different metallicities.  VMP acknowledges support from the Gemini Observatory.  SR acknowledges FAPESP.  HR acknowledges CNPq , CAPES and PROEX.}


\begin{thebibliography}{}
\bibitem[Abate et al.(2013)]{abate2013}Abate, C., Pols, O. R., Izzard, R. G., Mohamed, S. S., \& de Mink, S. E. 2013, \aap, 552, A26
\bibitem[Allende Prieto et al.(2008)]{allende2008}Allende Prieto, C., Sivarani, T., Beers, T. C., et al. 2008, \aj, 136, 2070
\bibitem[Aoki et al.(2007)]{aoki2007}Aoki, W., Beers, T. C., Christlieb, N., Norris, J. E., Ryan, S. G., \& Tsangarides, S. 2007, \apj, 655, 492
\bibitem[Asplund et al.(2009)]{asplund2009}Asplund, M., Grevesse, N., Sauval, A. J., \& Scott, P. 2009, \araa, 47, 481
\bibitem[Barcza \& Benk(2009)]{barcza2009}Barcza, S. \& Benk, J. M. 2009, AIP Conf. Proc., 1170, 250
\bibitem[Beers et al.(1992)]{beers1992}Beers, T. C., Preston, G. W., \& Shectman, S. A. 1992, \aj, 103, 1987
\bibitem[Beers \& Christlieb(2005)]{beers2005}Beers, T. C., \& Christlieb, N. 2005, \araa, 43, 531
\bibitem[Bono et al.(2007)]{bono2007}Bono, G., Caputo, F., \& Criscienzo, M. 2007, \aap, 476, 779
\bibitem[Cacciari et al.(2005)]{cacciari2005}Cacciari, C., Corwin, T., \& Carney, B. 2005, \aj, 129, 267
\bibitem[Campbell \& Lattanzio(2008)]{campbell2008}Campbell, S. W., \& Lattanzio, J. C. 2008, \aap, 490, 769
\bibitem[Carollo et al.(2012)]{carollo2012}Carollo, D., Beers, T. C., Bovy, J., et al. 2012, \apj, 744, 195
\bibitem[Castelli \& Kurucz(2004)]{castelli2004}Castelli, F., \& Kurucz, R. L. 2004, arXiv:astro-ph/0405087
\bibitem[Childress et al.(2013)]{childress2013}Childress, M. J., Vogt, F. P. A., Nielsen, J., \& Sharp, R. G. 2013, arXiv:1311.2666v1
\bibitem[Christlieb et al.(2008)]{christlieb2008}Christlieb, N., Sch\"{o}rck, T., Frebel, A., Beers, T. C., Wisotzki, L., \& Reimers, D. 2008, \aap, 484, 721
\bibitem[Cohen et al.(2005)]{cohen2005}Cohen, J. G., Shectman, S., Thompson, I., et al. 2005, \apj, 633, 109
\bibitem[Dopita et al.(2007)]{dopita2007}Dopita, M., Hart, J., McGregor, P., Oates, P., Bloxham, G., \& Jones, D. 2007, \apss, 310, 255
\bibitem[Frebel et al.(2006)]{frebel2006}Frebel, A., Christlieb, N., Norris, J. E., et al. 2006, \apj, 652, 1585
\bibitem[Hansen et al.(2011a)]{hansen2011}Hansen, C. J., Nordstr{\"o}m, B., Bonifacio, P., et al. 2011, \aap, 527, 65
\bibitem[Hansen et al.(2011b)]{hansent2011}Hansen, T., Andersen, J., Nordstr{\"o}m, B., Buchhave, L. A., \& Beers, T. C. 2011, \apj, 743, 1
\bibitem[Hansen et al.(2013)]{hansen2013}Hansen, T., Anderson, J., \& Nordstr{\"o}m, B. 2013, Proceedings of the XII International Symposium on Nuclei in the Cosmos (NIC XII), Proceedings of Science, 146, 193
\bibitem[Herwig(2005)]{herwig2005}Herwig, F. 2005, \araa, 43, 435
\bibitem[Hirschi et al.(2006)]{hirschi2006}Hirschi, R., Fr{\"o}hlich, C., Liebend{\"o}rfer, M., \& Thieleman, F.-K. 2006, Reviews of Modern Astronomy, 19, 101
\bibitem[Ito et al.(2013)]{ito2013}Ito, H., Aoki, W., Beers, T. C., Tominaga, N., Honda, S., \& Carollo, D. 2013, \apj, 733, 33 
\bibitem[Karakas et al.(2014)]{karakas2014}Karakas, A. I., Marino, A. F., \& Nataf, D. M. 2014, \apj, 784, 32
\bibitem[Kinemuchi et al.(2006)]{kinemuchi2006}Kinemuchi, K., Smith, H. A., Wozniak, P. R., \& McKay, T. A. 2006, \aj, 132, 1202
\bibitem[Kinman et al.(2012)]{kinman2012}Kinman, T. D., Aoki, W., Beers, T. C., \& Brown, W. R. 2012, \apjl, 755, L18
\bibitem[Kobayashi et al.(2011)]{kobayashi2011}Kobayashi, C., Tominaga, N., \& Nomoto, K. 2011, \apj, 730, L14 
\bibitem[Layden(1995)]{layden1995}Layden, A. 1995, \aj, 110, 2288
\bibitem[Lee et al.(2008a)]{lee2008a}Lee, Y. S., Beers, T. C., Sivarani, T., et al. 2008a, \aj, 136, 2022
\bibitem[Lee et al.(2008b)]{lee2008b}Lee, Y. S., Beers, T. C., Sivarani, T., et al. 2008b, \aj, 136, 2050
\bibitem[Lee et al.(2011)]{lee2011}Lee, Y. S., Beers, T. C., Allende Prieto, C., et al. 2011, \aj, 141, 90
\bibitem[Lee et al.(2013)]{lee2013}Lee, Y. S., Beers, T. C., Masseron, T., et al. 2013, \aj, 146, 132
\bibitem[Lucatello et al.(2006)]{lucatello2006}Lucatello, S., Beers, T. C., Christlieb, N., et al. 2006, \apj, 652, 37
\bibitem[Lugaro et al.(2012)]{lugaro2012}Lugaro, M., Karakas, A. I., Stancliffe, R. J., \& Rijs, C. 2012, \apj, 747, 2
\bibitem[Marsteller et al.(2005)]{marsteller2005}Marsteller, B., Beers, T. C., Rossi, S., Christlieb, N., Bessell, M., \& Rhee, J. 2005, Nucl. Phys. A, 758, 312
\bibitem[Masseron et al.(2012)]{masseron2012}Masseron, T., Johnson, J. A., Lucatello, S., et al. 2012, ApJ, 751, 14
\bibitem[Meynet et al.(2006)]{meynet2006}Meynet, G., Ekstr{\"o}m, S., \& Maeder, A. 2006, \aap, 447, 623
\bibitem[Meynet et al.(2010)]{meynet2010}Meynet, G.,Hirschi, R., Ekstr{\"o}m, S., et al. 2010, \aap, 521, 30
\bibitem[Miceli et al.(2008)]{miceli2008}Miceli, A., Rest, A., Stubbs, C., et al. 2008, \apj, 678, 865
\bibitem[Nomoto et al.(2013)]{nomoto2013}Nomoto, K., Kobayashi, C., \& Tominaga, N. 2013, \araa, 51, 457
\bibitem[Norris et al.(1997)]{norris1997}Norris, J. E., Ryan, S. G., \& Beers, T. C. 1997, \apj, 488, 350
\bibitem[Norris et al.(2007)]{norris2007}Norris, J. E., Christlieb, N., Korn, A. J., et al. 2007, \apj, 670, 774
\bibitem[Norris et al.(2013b)]{norris2013}Norris, J. E., Yong, D., Bessell, M. S., et al. 2013, \apj, 762, 28
\bibitem[Oosterhoff(1939)]{oosterhoff1939}Oosterhoff, P. T. 1939, The Observatory, 62, 104
\bibitem[Preston et al.(2006)]{preston2006}Preston, G. W., Thompson, I. A., Sneden, C., Stachoqski, G., \& Shectman, S. A. 2006, \aj, 132, 1714
\bibitem[Rossi et al.(2005)]{rossi2005}Rossi, S., Beers, T. C., Sneden, C., Sevastyanenko, T., Rhee, J., \& Marsteller, B. 2005, \aj, 130, 2804
\bibitem[Samus et al.(2009)]{samus2009}Samus, N. N., et al., General Catalog of Variable Stars, Vizier On-Line Data Catalog
\bibitem[Smith(2004)]{smith2004}Smith, H. A., RR Lyrae Stars, Cambridge Univ. Press (Cambridge)
\bibitem[Smolinski et al.(2011)]{smolinski2011}Smolinski, J. P., Lee, Y. S., Beers, T. C., et al. 2011, \aj, 141, 89
\bibitem[Sneden(1973)]{sneden1973}Sneden, C. A. 1973, PhD thesis, Univ. Texas at Austin
\bibitem[Sneden et al.(2008)]{sneden2008}Sneden, C., Cowan J. J., \& Gallino, R. 2008, \araa, 46, 241
\bibitem[Spite et al.(2013)]{spite2013}Spite, M., Caffau, E., Bonifacio, P., et al. 2013, \aap, 552, 107
\bibitem[Stancliffe et al.(2007)]{stancliffe2007}Stancliffe, R, J., Glebbeek, E., Izzard, R. G., \& Pols, O. R. 2007, \aap, 464, L57
\bibitem[Stancliffe \& Glebbeek(2008)]{stancliffe2008}Stancliffe R. J., \& Glebbeek, E. 2008, MNRAS, 389, 1828
\bibitem[Stancliffe(2009)]{stancliffe2009}Stancliffe, R. J. 2009, MNRAS, 394, 1051
\bibitem[Stancliffe et al.(2013)]{stancliffe2013}Stancliffe, R. J., Kennedy, C. R., Lau, H. H. B., \& Beers, T. C. 2013, \mnras, 435, 698
\bibitem[Tominaga et al.(2007)]{tominaga2007}Tominaga, N., Umeda, H., \& Nomoto, K. 2007, \apj, 660, 516
\bibitem[Tominaga et al.(2013)]{Tominaga2013}Tominaga, N., Iwamoto, N., \& Nomoto, K. 2013, \apj, in press (arXiv:1309.6734)
\bibitem[Umeda \& Nomoto(2003)]{umeda2003}Umeda, H., \& Nomoto, K. 2003, Nature, 422, 871
\bibitem[Vivas et al.(2004)]{vivas2004}Vivas, A. K., Zinn, R., Abad, C., et al. 2004, \aj, 127, 1158
\bibitem[Vivas et al.(2008)]{vivas2008}Vivas, A. K., Jaff{\'e}, Y. L., Zinn, R., Winnick, R., Duffau, S., \& Mateu, C. 2008, \aj, 136, 1645
\bibitem[Wallerstein et al.(2009)]{wallerstein2009}Wallerstein, G., Kovtyukh, V., \& Andreivsky, S. 2009, \apj, 692, L127
\end{thebibliography}
\end{document}